# Effects of distinct ion temperatures on the head-on collision and phase shifts of dust acoustic one and multi-solitons in dusty plasmas

M. S. Alam[1,3*], M. G. Hafez[1, 3**], M. R. Talukder[2] and M. Hossain Ali[1]

[1]Department of of Applied Mathematics, University of Rajshahi, Rajshahi-6205, Bangladesh.

[2]Plasma Science and Technology Lab, Department of Applied Physics and Electronic Engineering, University of Rajshahi, Rajshahi-6205, Bangladesh.

[3]Department of Mathematics, Chittagong University of Engineering and Technology, Chittagong-4349, Bangladesh.

*E-mail: alam21nov_2016@yahoo.com  ***E-mail: golam_hafez@yahoo.com

**Abstract:** The propagation characteristics and interactions between the dust acoustic (DA) one and multi-solitons in an unmagnetized dusty plasmas composing negatively charged mobile dust, Boltzmann distributed electrons, nonextensive distributed cold and nonthermal distributed hot ions are studied. The well known extended Poincaré-Lighthill-Kuo (PLK) method is employed to derive the two-sided Korteweg–de Vries (KdV) equations. The solutions of KdV equations are constructed using the Hirota's method both for one and multi solitons. The phase shifts are determined for the interaction of one, two and three DA solitons. The effects of plasma parameters on the head-on collision of DA one- as well as multi-solitons and their corresponding phase shifts are investigated.

## 1. Introduction

Dusty plasmas, composed of electron-ion plasma with an additional component of small micron or submicron sized extremely massive charged dust, are considered for understanding several types of collective processes that are existed in the lower and upper mesosphere, cometary tails, planetary rings, interstellar media, planetary magnetosphere, interplanetary spaces, etc [1-4] space as well as laboratory dusty plasmas [5-7]. The nonlinear collective effects of plasmas cannot appropriately be studied without tedious mathematical techniques. The localization of waves produces several types of important consistent structures, due to the interaction of nonlinearity with dispersion or dissipation, namely: solitary waves, shock waves, double layer (DL), vortices, etc which carries significant role both from the theoretical and experimental point of views. On the other hand, the plasmas significantly modify the wave spectra and introduce new eigen-modes [8, 9], such as dust-acoustic waves (DA), dust-ion acoustic waves (DIA), dust lattice waves, etc in the presence of charged dust components. Rao et al. [5] have investigated the characteristics of low-phase speed DA waves in dusty plasmas, which are existed both in the space and laboratory devices. It was found that the inertia provides by the mass of the dust particle and the pressure of the inertia-less electrons and ions provides restoring force due to the production of DA waves in the plasmas. Many authors [10-18] have investigated the propagation characteristics of DA waves in dusty plasmas considering various types of plasma assumptions. Tribeche and Merriche [19] have studied the effects of ion nonextensivity on DA solitary waves (DASWs). Yasmin et al. [20] have analyzed the properties of nonlinear waves in dusty plasmas consisting of $q$-distributed electrons. Sahu and Tribeche [21] have described the solitary and shock waves in dusty plasma associated with nonextensive ions. Besides, the face to face collision between the waves and their corresponding phase shifts also play a vital role to understand the physical phenomena in the plasmas. A few authors [22-27] have investigated the head-on collision of two solitary waves in distinct plasmas using the extended Poincaré-Lighthill-Kuo (PLK) method. Mandal et al. [18] have described the overtaking collision and phase shifts of DA multi-soliton in dusty plasmas with nonthermal electrons. Ghosh et al. [22] have investigated the head-on collision of DASWs in a four-component dusty plasmas with nonthermal ions.

Tasnim et al. [28] have studied the properties of DASWs in an unmagnetized quasi-neutral dusty plasma containing extremely massive negatively charged mobile dust particles, Boltzmann distributed electrons, and ions of two distinct temperatures, where the lower (cold) temperature ions follow nonextensive distribution, while higher (hot) temperature ions follow nonthermal distribution. They have derived the KdV, mKdV (modified KdV), and SG (standard Gardner) equations to study the nonlinear physical phenomena in the aforementioned plasmas. They have also showed that how the Gardner solution differs from the KdV and MKdV solutions and mentioned that the plasmas may exists in various cosmic dust-laden plasmas [29-31], where two-temperature ions [32-36] can significantly modify the wave dynamics. However, the role of the head-on collision between the DA waves may not be ignored, which also play important role for understanding physical scenarios in the aforementioned plasmas [28]. Being motivated, for the potentiality of the problems related to the astrophysical, space and laboratory plasmas, the head-on collisions between the DA one-, and multi-soliton, and their phase shifts in an unmagnetized plasmas consisting of extremely massive negatively charged mobile dust particles, Boltzmann distributed electrons, and ions of two distinct temperatures are investigated. The effects of cold and hot ions temperature ratio ($\sigma_1$), cold ion-electron temperature ratio ($\sigma_2$), unperturbed cold ion-dust density ratio ($\mu_{i1}$), unperturbed hot ion-dust density ratio ($\mu_{i2}$), strength of cold ion nenextensivity ($q$) and population of hot ion nonthermality ($\beta$) on the phase shift, and face to face collision between the DA one and multi-solitons are examined. Thus, the paper is organized as follows: theoretical model and derivations of two different types of KdV equations are depicted in Section 2. The one and multi-solitons solutions of KdV equations and their corresponding phase shifts are displayed in Section 3. The results along with relevant discussion are presented in Section 4. Finally, the conclusion is drawn in section 5.

## 2. Governing fluid and derivation of KdV equations

Let us consider a one dimensional unmagnatized plasma system consisting of negatively charged mobile dust, lower (cold) and higher (hot) temperature ions, and Boltzmann-distributed electrons. The lower and higher temperatures ions are assumed to follow the nonextensive and nonthermal distributions, respectively. At equilibrium, the charge neutrality condition is obtained as $n_{i10} + n_{i20} = n_{e0} + Z_d n_{d0}$, where $n_{i10}$ and $n_{i20}$ are the densities of unperturbed ions having lower and higher temperatures, $n_{e0}$ and $n_{d0}$ are the densities of unperturbed electrons and electrons on the surface of the dust grains, respectively, and $Z_d$ is the charge number. To study the face to face collision of DA waves and their corresponding phase shift in the plasmas, the normalized governing fluid equations can be defined [28] as

$$\frac{\partial n_d}{\partial t} + \frac{\partial}{\partial x}(n_d u_d) = 0 \qquad (1)$$

$$\frac{\partial u_d}{\partial t} + u_d \frac{\partial u_d}{\partial x} = \frac{\partial \phi}{\partial x} \qquad (2)$$

$$\frac{\partial^2 \phi}{\partial x^2} = n_d + \mu e^{\sigma_2 \phi} - \mu_{i1}[1-(q-1)\phi]^{\frac{(q+1)}{2(q-1)}} - \mu_{i2}(1+\beta\sigma_1\phi+\beta\sigma_1^2\phi^2)e^{-\sigma_1\phi}. \qquad (3)$$

Here, $n_d$ is dust particle density normalized by $n_{d0}$, $u_d$ is the dust fluid speed normalized by dust acoustic speed $C_d = (Z_d k_B T_{i1}/m_d)^{1/2}$, $\phi$ is the electrostatic potential normalized by $T_{i1}/e$, $\sigma_1 = T_{i1}/T_{i2}$, $\sigma_2 = T_{i1}/T_e$, $\mu_{i1} = n_{i10}/Z_d n_{d0}$, $\mu_{i2} = n_{i20}/Z_d n_{d0}$, $T_{i1}(T_{i2})$ is the lower (higher) ion temperature, $k_B$ is the Boltzmann constant and $m_d$ is the dust particle mass. The time variable $t$ is normalized by dust particle period $\omega_{pd}^{-1} = (m_d/4\pi n_{d0} Z_d^2 e^2)^{1/2}$ and the space variable $x$ is normalized by the Debye length $\lambda_{Dm} = (k_B T_{i1}/4\pi n_{d0} Z_d e^2)^{1/2}$. The equilibrium charge neutrality condition in the plasmas is obtained taking the Poisson's relation into account as $\mu = n_{e10}/Z_d n_{d0} = \mu_{i1} + \mu_{i2} - 1$.

To investigate the propagation characteristics and the interaction of the DA solitons, the scaling variables $x$ and $t$ can be stretched by the new coordinate system [23, 24] using the extended PLK method as

$$\left.\begin{aligned}\xi &= \varepsilon(x - Vt) + \varepsilon^2 P_0(\eta, \tau) + \varepsilon^3 P_1(\eta, \xi, \tau) + \ldots \ldots \\ \eta &= \varepsilon(x + Vt) + \varepsilon^2 Q_0(\xi, \tau) + \varepsilon^3 Q_1(\eta, \xi, \tau) + \ldots \ldots \\ \tau &= \varepsilon^3 t\end{aligned}\right\}, \tag{4}$$

where $\xi$ and $\eta$ are the trajectories between the solitons which are traveling toward each other, and $V$ is the unknown phase velocity of DA waves and $\varepsilon$ is a small parameter measuring the strength of nonlinearity and dissipation. The unknown variables $P_0(\eta, \tau)$ and $Q_0(\xi, \tau)$ will be determined later. Using Eq.(4), the operators can be defined as

$$\left.\begin{aligned}\frac{\partial}{\partial t} &\approx \varepsilon^3 \frac{\partial}{\partial \tau} + \varepsilon V\left(-\frac{\partial}{\partial \xi} + \frac{\partial}{\partial \eta}\right) + \varepsilon^3 V\left(\frac{\partial P_0}{\partial \eta}\frac{\partial}{\partial \xi} - \frac{\partial Q_0}{\partial \xi}\frac{\partial}{\partial \eta}\right) + \ldots \ldots \\ \frac{\partial}{\partial x} &\approx \varepsilon\left(\frac{\partial}{\partial \xi} + \frac{\partial}{\partial \eta}\right) + \varepsilon^3\left(\frac{\partial P_0}{\partial \eta}\frac{\partial}{\partial \xi} + \frac{\partial Q_0}{\partial \xi}\frac{\partial}{\partial \eta}\right) + \ldots \ldots \ldots \ldots\end{aligned}\right\}. \tag{5}$$

The perturbed quantities can be expand in power series of $\varepsilon$ as

$$\begin{bmatrix}n_d \\ u_d \\ \phi\end{bmatrix} = \begin{bmatrix}1 \\ 0 \\ 0\end{bmatrix} + \varepsilon^2 \begin{bmatrix}n_d^{(1)} \\ u_d^{(1)} \\ \phi^{(1)}\end{bmatrix} + \varepsilon^3 \begin{bmatrix}n_d^{(2)} \\ u_d^{(2)} \\ \phi^{(2)}\end{bmatrix} + \varepsilon^4 \begin{bmatrix}n_d^{(3)} \\ u_d^{(3)} \\ \phi^{(3)}\end{bmatrix} + \cdots \tag{6}$$

Inserting Eqs. (5) and (6) into Eqs. (1)-(3) and equating the quantities with equal power of $\varepsilon$, one may obtain a set of equations in different orders of $\varepsilon$. To the lowest order of $\varepsilon$ yeild

$$\left(-V\frac{\partial n_d^{(1)}}{\partial \xi} + \frac{\partial u_d^{(1)}}{\partial \xi}\right) + \left(V\frac{\partial n_d^{(1)}}{\partial \eta} + \frac{\partial u_d^{(1)}}{\partial \eta}\right) = 0, \tag{7}$$

$$\left(-V\frac{\partial u_d^{(1)}}{\partial \xi} - \frac{\partial \phi^{(1)}}{\partial \xi}\right) + \left(V\frac{\partial u_d^{(1)}}{\partial \eta} - \frac{\partial \phi^{(1)}}{\partial \eta}\right) = 0, \tag{8}$$

$$n_d^{(1)} = -C_1 \phi^{(1)}, \tag{9}$$

where $C_1 = \left\{\mu\sigma_2 + \frac{1}{2}\mu_{i1}(q+1) - \mu_{i2}\sigma_1(\beta-1)\right\}$. One may define the relations along with the different physical quantities taking Eqs. (7)-(9) into account as

$$\phi^{(1)} = \phi_\xi^{(1)}(\xi,\tau) + \phi_\eta^{(1)}(\eta,\tau), \tag{10}$$

$$n_d^{(1)} = -C_1\left[\phi_\xi^{(1)}(\xi,\tau) + \phi_\eta^{(1)}(\eta,\tau)\right], \tag{11}$$

$$u_d^{(1)} = \frac{1}{V}\left[-\phi_\xi^{(1)}(\xi,\tau) + \phi_\eta^{(1)}(\eta,\tau)\right]. \tag{12}$$

The normalized phase velocity is obtained [28] as $V = \left\{\mu\sigma_2 + \frac{1}{2}\mu_{i1}(q+1) - \mu_{i2}\sigma_1(\beta-1)\right\}^{-\frac{1}{2}}$ taking the solvability condition into account. The considered functions $\phi_\xi^{(1)}(\xi,\tau)$ and $\phi_\eta^{(1)}(\eta,\tau)$ may be determined taking the next order of $\varepsilon$. From the relations of Eqs. (10)-(12), the two sided electrostatic waves may appear, one of which $(\phi_\xi^{(1)})$ is traveling to right direction and the other $(\phi_\eta^{(1)})$ is traveling to left direction. To the next order of $\varepsilon$, one can obtain another set of equations whose solutions are defined as

$$\phi^{(2)} = \phi_\xi^{(2)}(\xi,\tau) + \phi_\eta^{(2)}(\eta,\tau), \tag{13}$$

$$n_d^{(2)} = -C_1\left[\phi_\xi^{(2)}(\xi,\tau) + \phi_\eta^{(2)}(\eta,\tau)\right], \tag{14}$$

$$u_d^{(2)} = \frac{1}{V}\left[-\phi_\xi^{(2)}(\xi,\tau) + \phi_\eta^{(2)}(\eta,\tau)\right]. \tag{15}$$

Finally, the next higher order of $\varepsilon$ gives

$$\frac{\partial n_d^{(1)}}{\partial \tau} - V\frac{\partial n_d^{(3)}}{\partial \xi} + V\frac{\partial n_d^{(3)}}{\partial \eta} + \frac{\partial u_d^{(3)}}{\partial \xi} + \frac{\partial u_d^{(3)}}{\partial \eta} + \frac{\partial}{\partial \xi}\left(n_d^{(1)}u_d^{(1)}\right) + \frac{\partial}{\partial \eta}\left(n_d^{(1)}u_d^{(1)}\right)$$

$$+ V\frac{\partial P_0}{\partial \eta}\frac{\partial n_d^{(1)}}{\partial \xi} + \frac{\partial P_0}{\partial \eta}\frac{\partial u_d^{(1)}}{\partial \xi} - V\frac{\partial Q_0}{\partial \xi}\frac{\partial n_d^{(1)}}{\partial \eta} + \frac{\partial Q_0}{\partial \xi}\frac{\partial u_d^{(1)}}{\partial \eta} = 0, \tag{16}$$

$$\frac{\partial u_d^{(1)}}{\partial \tau} - V\frac{\partial u_d^{(3)}}{\partial \xi} + V\frac{\partial u_d^{(3)}}{\partial \eta} + u_d^{(1)}\frac{\partial u_d^{(1)}}{\partial \xi} + u_d^{(1)}\frac{\partial u_d^{(1)}}{\partial \eta} - \frac{\partial \phi^{(3)}}{\partial \xi} - \frac{\partial \phi^{(3)}}{\partial \eta} + V\frac{\partial P_0}{\partial \eta}\frac{\partial u_d^{(1)}}{\partial \xi}$$

$$- \frac{\partial P_0}{\partial \eta}\frac{\partial \phi^{(1)}}{\partial \xi} - V\frac{\partial Q_0}{\partial \xi}\frac{\partial u_d^{(1)}}{\partial \eta} - \frac{\partial Q_0}{\partial \xi}\frac{\partial \phi^{(1)}}{\partial \eta} = 0, \tag{17}$$

$$\frac{\partial^2 \phi^{(1)}}{\partial \xi^2} + \frac{\partial^2 \phi^{(1)}}{\partial \eta^2} + 2\frac{\partial^2 \phi^{(1)}}{\partial \xi \partial \eta} = n_d^{(3)} + C_1 \phi^{(3)} + C_2 \{\phi^{(1)}\}^2. \tag{18}$$

Simplifying Eqs. (16)-(18) and then integrating with regards to $\xi$ and $\eta$ provides

$$\begin{aligned}
2V^2 u_d^{(3)} =& \int \left( \frac{\partial \phi_\xi^{(1)}}{\partial \tau} + A\phi^{(1)} \frac{\partial \phi^{(1)}}{\partial \xi} + B \frac{\partial^3 \phi^{(1)}}{\partial \xi^3} \right) d\eta \\
&+ \int \left( \frac{\partial \phi_\eta^{(1)}}{\partial \tau} - A\phi_\eta^{(1)} \frac{\partial \phi_\eta^{(1)}}{\partial \eta} - B \frac{\partial^3 \phi_\eta^{(1)}}{\partial \eta^3} \right) d\xi \\
&+ \iint \left( 2V \frac{\partial P_0}{\partial \eta} - \left[ C_2 V^3 - \frac{1}{2V} \right] \phi_\eta^{(1)} \right) \frac{\partial^2 \phi_\xi^{(1)}}{\partial \xi^2} d\xi d\eta \\
&- \iint \left( 2V \frac{\partial Q_0}{\partial \xi} - \left[ C_2 V^3 - \frac{1}{2V} \right] \phi_\xi^{(1)} \right) \frac{\partial^2 \phi_\eta^{(1)}}{\partial \eta^2} d\xi d\eta,
\end{aligned} \tag{19}$$

where $A = -[(3/2V) + V^3 C_2]$ and $B = V^3/2$. The first and second term in the right side of (19) are proportional to $\eta$ and $\xi$, respectively, because the integrand functions involving in first and second terms in the right side of Eq.(19) are independent of $\eta$ and $\xi$, respectively. All the term of the first two expressions in the right side of Eq.(19) becomes secular. It may eliminate in order to stay away from specious resonances. Hence, one can derive the following KdV equations:

$$\frac{\partial \phi_\xi^{(1)}}{\partial \tau} + A \phi_\xi^{(1)} \frac{\partial \phi_\xi^{(1)}}{\partial \xi} + B \frac{\partial^3 \phi_\xi^{(1)}}{\partial \xi^3} = 0, \tag{20}$$

$$\frac{\partial \phi_\eta^{(1)}}{\partial \tau} - A \phi_\eta^{(1)} \frac{\partial \phi_\eta^{(1)}}{\partial \eta} - B \frac{\partial^3 \phi_\eta^{(1)}}{\partial \eta^3} = 0. \tag{21}$$

It is seen from Eqs. (20) and (21) that they yield two sided traveling wave KdV equations in the considered frame of references $\xi$ and $\eta$, respectively. Moreover, the third and fourth terms in the right side of Eq. (19) may become secular terms in the next higher order and yields the following equations:

$$\frac{\partial P_0}{\partial \eta} = D \phi_\eta^{(1)}, \tag{22}$$

$$\frac{\partial Q_0}{\partial \xi} = D \phi_\xi^{(1)}, \tag{23}$$

where,

$$D = \left[\frac{C_2 V^2}{2} - \frac{1}{4V^2}\right]. \tag{24}$$

The variables $P_0(\eta, \tau)$ and $Q_0(\xi, \tau)$ can be obtained by solving Eqs. (22) and (23) with the help of analytical solutions of the KdV equations (20) and (21).

## 3. Soliton solutions and phase shifts

To study the nonlinear propagation of face to face one as well multi-soliton collisions and their phase shifts of DA waves in the plasmas, one may derive the analytical soliton solutions of the KdV Eqs. (20) and (21). The Hirota bilinear method [37] is a well establish method for the determination of one and multiple-soliton solutions of the nonlinear partial differential equations. The one-soliton solutions of the KdV equations (20) and (21) can be written as

$$\phi_\xi^{(1)} = \frac{12B}{A} \frac{\partial^2}{\partial \xi^2} \left[\ln\left\{1 + \exp\left(k_1 B^{-\frac{1}{3}}\xi - k_1^3 \tau\right)\right\}\right], \tag{25}$$

$$\phi_\eta^{(1)} = \frac{12B}{A} \frac{\partial^2}{\partial \eta^2} \left[\ln\left\{1 + \exp\left(-k_1 B^{-\frac{1}{3}}\eta - k_1^3 \tau\right)\right\}\right]. \tag{26}$$

Using (25) and (26), Eqs. (22) and (23) can be converted to

$$\frac{\partial P_0}{\partial \eta} = \frac{12BD}{A} \frac{\partial^2}{\partial \eta^2} \left[\ln\{1 + \exp(-k_1 B^{-1/3}\eta - k_1^3 \tau)\}\right], \tag{27}$$

$$\frac{\partial Q_0}{\partial \xi} = \frac{12BD}{A} \frac{\partial^2}{\partial \xi^2} \left[\ln\{1 + \exp(k_1 B^{-1/3}\xi - k_1^3 \tau)\}\right]. \tag{28}$$

Solving Eqs.(27) and (28), the leading phase changes due to the collisions of two sided solitary waves can be obtained as

$$P_0(\eta, \tau) = -\frac{12B^{2/3} D k_1}{A} \frac{\exp(-k_1 B^{-1/3}\eta - k_1^3 \tau)}{1 + \exp(-k_1 B^{-1/3}\eta - k_1^3 \tau)}, \tag{29}$$

$$Q_0(\xi, \tau) = \frac{12B^{2/3} D k_1}{A} \frac{\exp\left(k_1 B^{-\frac{1}{3}}\xi - k_1^3 \tau\right)}{1 + \exp\left(k_1 B^{-\frac{1}{3}}\xi - k_1^3 \tau\right)}. \tag{30}$$

The trajectories of two solitary waves for weak face to face collision can be written as

$$\xi = \varepsilon(x - Vt) - \varepsilon^2 \frac{12B^{2/3}Dk_1}{A} \frac{\exp(-k_1 B^{-1/3}\eta - k_1^3\tau)}{1 + \exp(-k_1 B^{-1/3}\eta - k_1^3\tau)} + \cdots, \qquad (31)$$

$$\eta = \varepsilon(x + Vt) + \varepsilon^2 \frac{12B^{2/3}Dk_1}{A} \frac{\exp\left(k_1 B^{-\frac{1}{3}}\xi - k_1^3\tau\right)}{1 + \exp\left(k_1 B^{-\frac{1}{3}}\xi - k_1^3\tau\right)} + \cdots. \qquad (32)$$

To evaluate the phase shifts after a head-on collision of the two solitons, one may consider the solitons, say, $S_1$ and $S_2$ are, asymptotically, far from each other at the initial time. After collision, $S_1$ is far to the right of $S_2$. Using the relation $\nabla P_0 = \varepsilon(x - Vt)|_{\eta \to -\infty, \xi=0} - \varepsilon(x - Vt)|_{\eta \to \infty, \xi=0}$ and $\nabla Q_0 = \varepsilon(x + Vt)|_{\xi \to -\infty, \eta=0} - \varepsilon(x + Vt)|_{\xi \to \infty, \eta=0}$, the corresponding phase shifts may obtain as

$$\nabla P_0 = -\varepsilon^2 \frac{12B^{2/3}D}{A} k_1, \qquad (33)$$

$$\nabla Q_0 = \varepsilon^2 \frac{12B^{2/3}D}{A} k_1. \qquad (34)$$

Again, two-soliton solutions of Eqs. (20) and (21) can be written as

$$\phi_\xi^{(1)} = \frac{12B}{A} \frac{\partial^2}{\partial \xi^2} [\ln\{1 + \exp(\vartheta_1) + \exp(\vartheta_2) + a_{12} \exp(\vartheta_1 + \vartheta_2)\}], \qquad (35)$$

$$\phi_\eta^{(1)} = \frac{12B}{A} \frac{\partial^2}{\partial \eta^2} [\ln\{1 + \exp(\Omega_1) + \exp(\Omega_2) + a_{12} \exp(\Omega_1 + \Omega_2)\}], \qquad (36)$$

where $\vartheta_i = k_i B^{-1/3}\xi - k_i^3 \tau$, $\Omega_i = -k_i B^{-1/3}\eta - k_i^3 \tau$ and $a_{12} = (k_2 - k_1)^2/(k_2 + k_1)^2$ with $i = 1, 2$. Using Eqs.(35) and (36), the solution of Eqs. (22) and (23) can determine as

$$P_0 = -\frac{12B^{2/3}D}{A} \frac{k_1 \exp(\vartheta_1) + k_2 \exp(\vartheta_2) + a_{12}(k_1 + k_2)\exp(\vartheta_1 + \vartheta_2)}{1 + \exp(\vartheta_1) + \exp(\vartheta_2) + a_{12}\exp(\vartheta_1 + \vartheta_2)}, \qquad (37)$$

$$Q_0 = \frac{12B^{2/3}D}{A} \frac{k_1 \exp(\Omega_1) + k_2 \exp(\Omega_2) + a_{12}(k_1 + k_2)\exp(\Omega_1 + \Omega_2)}{1 + \exp(\Omega_1) + \exp(\Omega_2) + a_{12}\exp(\Omega_1 + \Omega_2)}, \qquad (38)$$

and the corresponding phase shifts may be obtained as

$$\nabla P_0 = -\varepsilon^2 \frac{12B^{2/3}D}{A}(k_1 + k_2), \qquad (39)$$

$$\nabla Q_0 = \varepsilon^2 \frac{12B^{2/3}D}{A}(k_1 + k_2). \tag{40}$$

Finally, three-soliton solutions of the Eqs. (20) and (21) can be written as

$$\phi_\xi^{(1)} = \frac{12B}{A}\frac{\partial^2}{\partial \xi^2}[\ln\{1 + \exp(\vartheta_1) + \exp(\vartheta_2) + \exp(\vartheta_3) + a_{12}\exp(\vartheta_1 + \vartheta_2)$$
$$+ a_{23}\exp(\vartheta_2 + \vartheta_3) + a_{13}\exp(\vartheta_1 + \vartheta_3)$$
$$+ a_{123}\exp(\vartheta_1 + \vartheta_2 + \vartheta_3)\}], \tag{41}$$

$$\phi_\eta^{(1)} = \frac{12B}{A}\frac{\partial^2}{\partial \eta^2}[\ln\{1 + \exp(\Omega_1) + \exp(\Omega_2) + \exp(\Omega_3) + a_{12}\exp(\Omega_1 + \Omega_2)$$
$$+ a_{23}\exp(\Omega_2 + \Omega_3) + a_{13}\exp(\Omega_1 + \Omega_3)$$
$$+ a_{123}\exp(\Omega_1 + \Omega_2 + \Omega_3)\}], \tag{42}$$

where $\vartheta_i = k_i B^{-1/3}\xi - k_i^3\tau$, $\Omega_i = -k_i B^{-1/3}\eta - k_i^3\tau$, $i = 1-3$, $a_{12} = (k_1 - k_2)^2/(k_1 + k_2)^2$, $a_{23} = (k_2 - k_3)^2/(k_2 + k_3)^2$, $a_{13} = (k_1 - k_3)^2/(k_1 + k_3)^2$ and $a_{123} = a_{12}a_{23}a_{13}$.

and their phase shifts may be evaluated as

$$\nabla P_0 = -\varepsilon^2 \frac{12B^{2/3}D}{A}(k_1 + k_2 + k_3), \tag{43}$$

$$\nabla Q_0 = \varepsilon^2 \frac{12B^{2/3}D}{A}(k_1 + k_2 + k_3). \tag{44}$$

The interactions among the DA solitons and their corresponding phase shifts on plasma parameters are discussed in the next section.

## 4. Results and Discussion

The head-on collision phenomena between the DA solitons and their corresponding phase shifts have investigated by deriving two-sided KdV equations involving nonlinearity ($A$) and dispersion ($B$) cofficients using the extended PLK method in the considered plasmas. The coefficients $A$ and $B$ strongly depends on the plasma parameters $\sigma_1 = T_{i1}/T_{i2}$, $\sigma_2 = T_{i1}/T_e$, $\mu_{i1} = n_{i10}/Z_d n_{d0}$, $\mu_{i2} = n_{i20}/Z_d n_{d0}$, $q$ and $\beta$. It is seen that the compressive and rarefactive DA solitons may exists when $A > 0$ and $A < 0$, respectively. Tasnim et al. [28] have clearly indicated that the small amplitude compressive, rarefactive and no DASWs exists when $\mu_{i1} < 0.56$, $\mu_{i1} > 0.56$ and $\mu_{i1} \sim 0.56$, respectively. On the other hand, Ghosh et al. [22] have mentioned that the positive or negative phase shift does not depend on the type of wave mode, but depends on the co-efficient $D$ in Eq. (26). Furthermore, many authors [26, 37, 38] have shown that the colliding acoustic wave has a positive phase shift in its traveling direction. El-Labany et al. [27] have illustrated that the colliding DASW has a negative phase shift in its traveling direction. Therefore, the phase shifts becomes either positive or negative due to head-on collision of the solitons. It is found that the positive phase shifts are obtained only if $A > 0$ and $D < 0$, otherwise there will be negative phase shifts. The parametric effects considering the typical data of Tasnim et al. [28] on the nonlinear head-on collision of electrostatic DA one and multi-solitons and their corresponding phase shifts are discussed.

Figures 1(a)-(c) show the effects on phase shift $\nabla P_0$ for the interaction between the two same amplitudes one-soliton with $\mu_{i1}$ and $\mu_{i2}$, $\sigma_1$ and $\sigma_2$, and $q$ and $\beta$, respectively taking the remaining parameters constant. On the other hand, Figures 2(a)-(c) show the changes of phase shifts $\nabla P_0$ for the interaction between the two same amplitudes with $\mu_{i1}$ and $\mu_{i2}$, $\sigma_1$ and $\sigma_2$, and $q$ and $\beta$, respectively considering the same values as in Fig. 1 except the wave numbers $k_1 = 1$ and $k_2 = 2$. It is seen that the phase shifts of head-on collision of two-sided one and multi solitons are strongly depend on the plasma parameters and the wave numbers, and are increasing with increasing $\mu_{i1}$, $\mu_{i2}$, $\sigma_1$ and $q$, and are decreasing with increasing $\sigma_2$ and $\beta$. It is also found the phase shifts for head-on two-soliton is higher rather than one-soliton due to the increase of wave number.

Figures 3(a)-(d) display the electrostatic potential structures $\phi_\xi^{(1)}$ and $\phi_\eta^{(1)}$ for one-soliton as mentioned in Eqs. (25) and (26) against $\xi$ and $\eta$, respectively considering the different values of the remaining plasma parameters and time $\tau$. It is seen that the amplitudes due to head-on collision of DASWs is increasing with the increase of $\mu_{i1}$, $\mu_{i2}$ and $\sigma_1$, and is decreasing with the increase of $\sigma_2$. It is observed that the coefficients of nonlinear term (A) of the two-sided KdV equations are decreasing with the increase of $\mu_{i1}$, $\mu_{i2}$ and $\sigma_1$. Figures 3 show the phenomena of head-on collisions of the one-solitons. It is found that Fig. 3(d) is the mirror image of Fig. 3(b), as is expected. It is also seen from Fig. 3(a) that the compressive and rarefactive potential structures are found for $\mu_{i1} < 0.56$ and $\mu_{i1} > 0.56$, respectively due to the interaction between two-solitons which are in good agreement with the investigations of [28]. Figures 4 display the electrostatic potential structures $\phi_\xi^{(1)}$ and $\phi_\eta^{(1)}$ for two-solitons as obtained from Eqs. (35) and (36) against $\xi$ and $\eta$, respectively, taking the different values

of the remaining plasma parameters and time $\tau$. It is seen that the amplitude of DA multi solitons is increasing with the increase of $\mu_{i1}$, $\mu_{i2}$ and $\sigma_1$, and is decreasing with the increase of $\sigma_2$ and $\beta$. Figures 4 clearly dictate that the four, compressive for $\mu_{i1} < 0.56$ and rarefactive for $\mu_{i1} > 0.56$, scattered solitons are produced due to the head-on collisions of two-solitons, in which of two are propagating from left to right and remaining two are propagating in the opposite direction. Finally, Figures 5 display the electrostatic potential structures $\phi_\xi^{(1)}$ and $\phi_\eta^{(1)}$ for three-soliton as obtained from Eqs. (41) and (42) against $\xi$ and $\eta$, respectively considering different values of the remaining plasma parameters and $\tau$. It is also seen that the amplitude of IASWs is increasing with the increase of $\mu_{i1}$, $\mu_{i2}$ and $\sigma_1$, and is decreasing with the increase of $\sigma_2$ and $\beta$. Figures 5 clearly indicate that the six, compressive for $\mu_{i1} < 0.56$ and rarefactive for $\mu_{i1} > 0.56$, scattered solitons are produced due to the head-on collisions of three-solitons, in which of three are propagating from left to right and remaining three are propagating in the opposite direction. Thus, the results obtained in this manuscript may also useful for better understanding the physical phenomena observed in various cosmic dust-laden plasmas.

## 5. Summary

The propagation characteristics and the interaction between the DA solitons composing negatively charged mobile dust, Boltzmann-distributed electrons and two distinct temperatures of ions are investigated. The KdV equations are derived using the extended PLK method. The analytical solutions for solitons are constructed using the well established Hirota bilinear method. The phase shifts due to head-on collisions among the DA one, two, and three solitons are determined analytically from the two-sided KdV equations. The effects of plasma parameters on the head-on collision between the electrostatic DA one and multi solitons and their corresponding phase shifts are discussed. The compressive and rarefactive scattering for two, four and six DA waves are observed for $\mu_{i1} < 0.56$ and $\mu_{i1} > 0.56$, respectively. The phase shifts due to head-on collisions of DA one as well as multi solitons are strongly depend on the plasma parameters and the wave numbers, and are increasing with increasing $\mu_{i1}$, $\mu_{i2}$, $\sigma_1$ and $q$, and are decreasing with the increase of $\sigma_2$ and $\beta$. One may conclude that the results obtained in this investigation might be useful for understanding electrostatic disturbances in some space and laboratory plasma systems, such as Saturn's Ering Saturn's F-ring, noctilucent clouds, Halley's comet, interstellar molecular clouds in cosmic dust-laden plasma, laboratory dusty plasmas, etc., where major plasma species are negatively charged mobile dust, Boltzmann distributed electrons, and two temperature ions following nonextensive and nonthermal distributions.

**References**

1. P. K. Shukla and A. A. Mamun, Introduction to Dusty Plasma Physics (IOP, Bristol, 2002).
2. P. K. Shukla, Phys. Plasmas **8**, 1791 (2001).
3. D. A. Mendis and M. Rosenberg, Annu. Rev. Astron. Astrophys. **32**, 418 (1994).
4. F. Verheest, Waves in Dusty Plasmas (Kluwer, Dordrecht, 2000).
5. N. N. Rao, P. K. Shukla, and M. Y. Yu, Planet. Space Sci. **38**, 543 (1990).
6. A. Barkan, R. L. Merlino, and N. D'Angelo, Phys. Plasmas **2**, 3563 (1995).
7. J. T. Mendonça, N. N. Rao, and A. Guerreiro, Euro-phys. Lett. **54**, 741 (2001).
8. Y. Nakamura, H. Bailung, and P. K. Shukla, Phys. Rev. Lett. **83**, 1602 (1999).
9. R. L. Merlino and J. Goree, Phys. Today **57**, 32 (2004).
10. A. A. Mamun, R. A. Cairns, and P. K. Shukla, Phys. Plasmas **3**, 702 (1996).
11. A. A. Mamun, Astrophys. Space Sci. **260**, 507 (1998).
12. A. A. Mamun, S. M. Russel, C. A. M. Briceno, M. N. Alan, T. K. Datta, and A. K. Das, Planet. Space Sci. **48**, 599 (2000).
13. C. A. Mendoza-Briceno, S. M. Russel, and A. A. Mamun, Planet. Space Sci. **48**, 599 (2000).
14. W. S. Duan, H. Y. Wang, and P. John, Commun. Theor. Phys. **45**, 1112 (2006).
15. M. M. Lin and W. S. Duan, Chaos Solitons Fractals **33**, 1189 (2007).
16. H. Y. Wang and W. S. Duan, Acta Phys. Sin. **56**, 3977 (2007).
17. H. Y. Wang and K. B. Zhang, Can. J. Phys. **86**, 1381(2008).
18. G. Mandal, K. Roy, A. Paul, A. Saha and P. Chatterjee, Z. Naturforsch.70(9)a, 703 (2015).
19. M. Tribeche and A. Merriche, Phys. Plasmas **18**, 034502 (2011).
20. S. Yasmin, M. Asaduzzaman, and A. A. Mamun, Astrophys. Space Sci. **343**, 245 (2012).
21. B. Sahu and M. Tribeche, Astrophys. Space Sci. **338**, 259 (2012).
22. U. N. Ghosh, K. Roy and P. Chatterjee, Phys. Plasmas **18**, 103703 (2011).
23. S. Parveen, S. Mahmood, M. Adnan and A. Qamar, Phys. Plasmas **23**, 092122 (2016).
24. K. Roy, M. K. Ghorui, P. Chatterjee and M. Tribeche, Commun. Theor. Phys. **65**, 237(2016).
25. J. N. Han, S. C. Li, X. X. Yang, and W. S. Duan, Eur. Phys. J. D **47**, 197 (2008).
26. J. N. Han, S. L. Du, and W. S. Duan, Phys. Plasmas 15, 112104 (2008).
27. S. K. El-Labany, E. F. El-Shamy, R. Sabry, and M. Shokry, Astrophys. Space Sci. **325**, 201 (2010).
28. I. Tasnim , M. M. Masud and A. A. Mamun, Plasma Phys. Reports **40**, 723(2014).
29. S. Imada, H. Hara, and T. Watanabe, Astrophys. J. Lett. **705**, L208 (2009).
30. M. Tribeche and A. Merriche, Phys. Plasmas **18**, 034502 (2011).
31. K. B. Zhang and H. Y. Wang, J. Kor. Phys. Soc. **55**, 1461 (2009).
32. I. Tasnim, M. M. Masud, and A. A. Mamun, Astrophys. Space Sci. **343**, 647 (2013).

33. M. M. Masud and A. A. Mamun, JETP Lett. **96**, 855 (2012).
34. M. Asaduzzaman and A. A. Mamun, Astrophys. Space Sci. **341**, 535 (2012).
35. I. Tasnim, M. M. Masud, M. Asaduzzaman, and A. A. Mamun, Chaos **23**, 013147 (2013).
36. D. Dorranian and A. Sabetkar, Phys. Plasmas **19**, 013702 (2012).
37. J. K. Xue, Phys. Rev. E 69, 016403 (2004).
38. G. Z. Liang, J. N. Han, M. M. Lin, J. N. Wei, and W. S. Duan, Phys. Plasmas 16, 073705 (2009).
39. R. Hirota, The Direct Method in The Soliton Theory (Cambridge University Press, Cambridge, 2004).

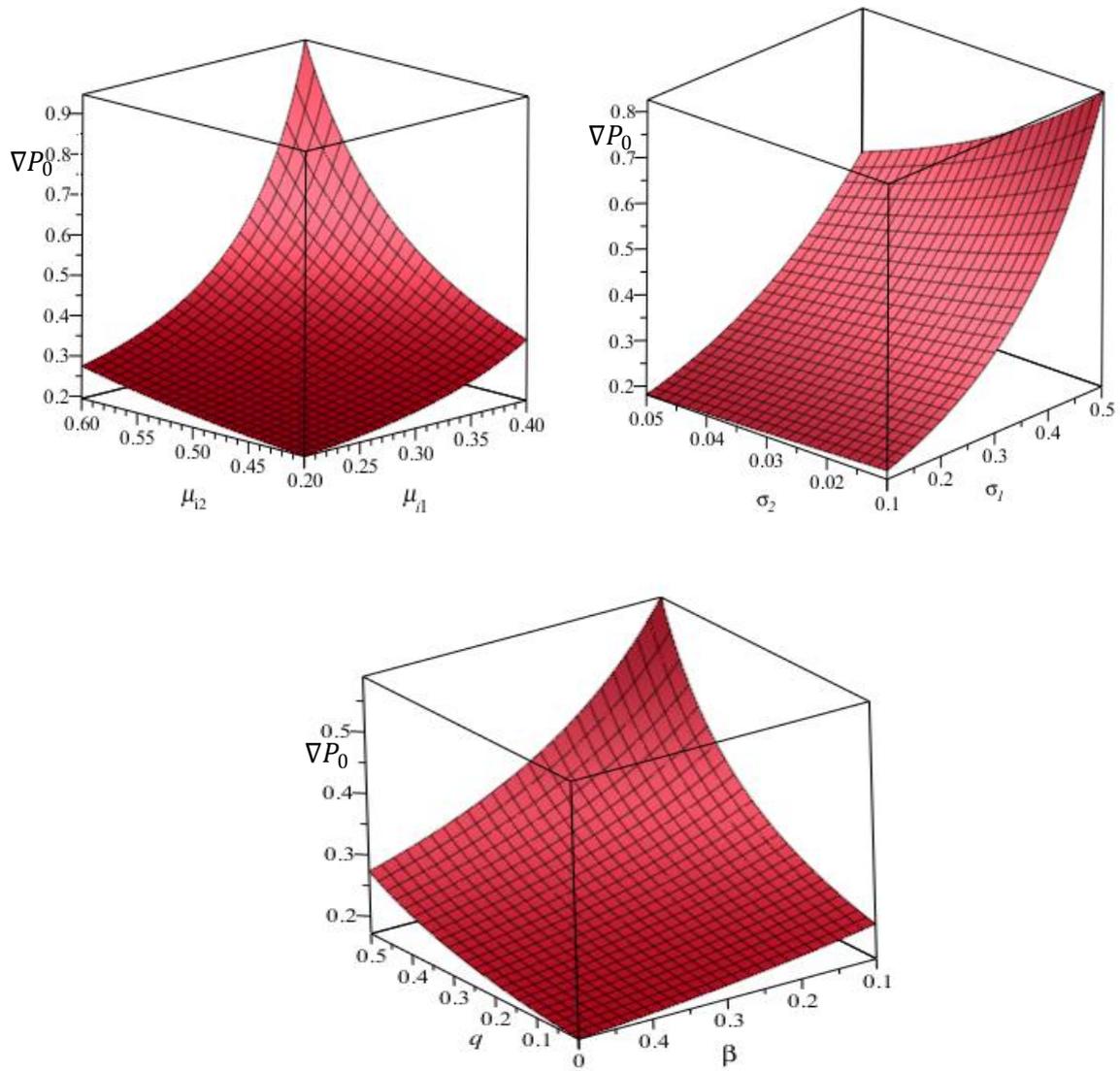

Fig. 1 Influence on phase shifts due to head-on collision for one-soliton (a) $\mu_{i1}$ and $\mu_{i2}$ taking $\sigma_1 = 0.2$, $\sigma_2 = 0.03$, $\beta = 0.3$ and $q = 0.25$, (b) $\sigma_1$ and $\sigma_2$ taking $\mu_{i1} = 0.3$, $\mu_{i2} = 0.41$, $\beta = 0.3$ and $q = 0.25$, and (c) $q$ and $\beta$ taking $\sigma_1 = 0.2$, $\sigma_2 = 0.03$, $\mu_{i1} = 0.3$, $\mu_{i2} = 0.41$.

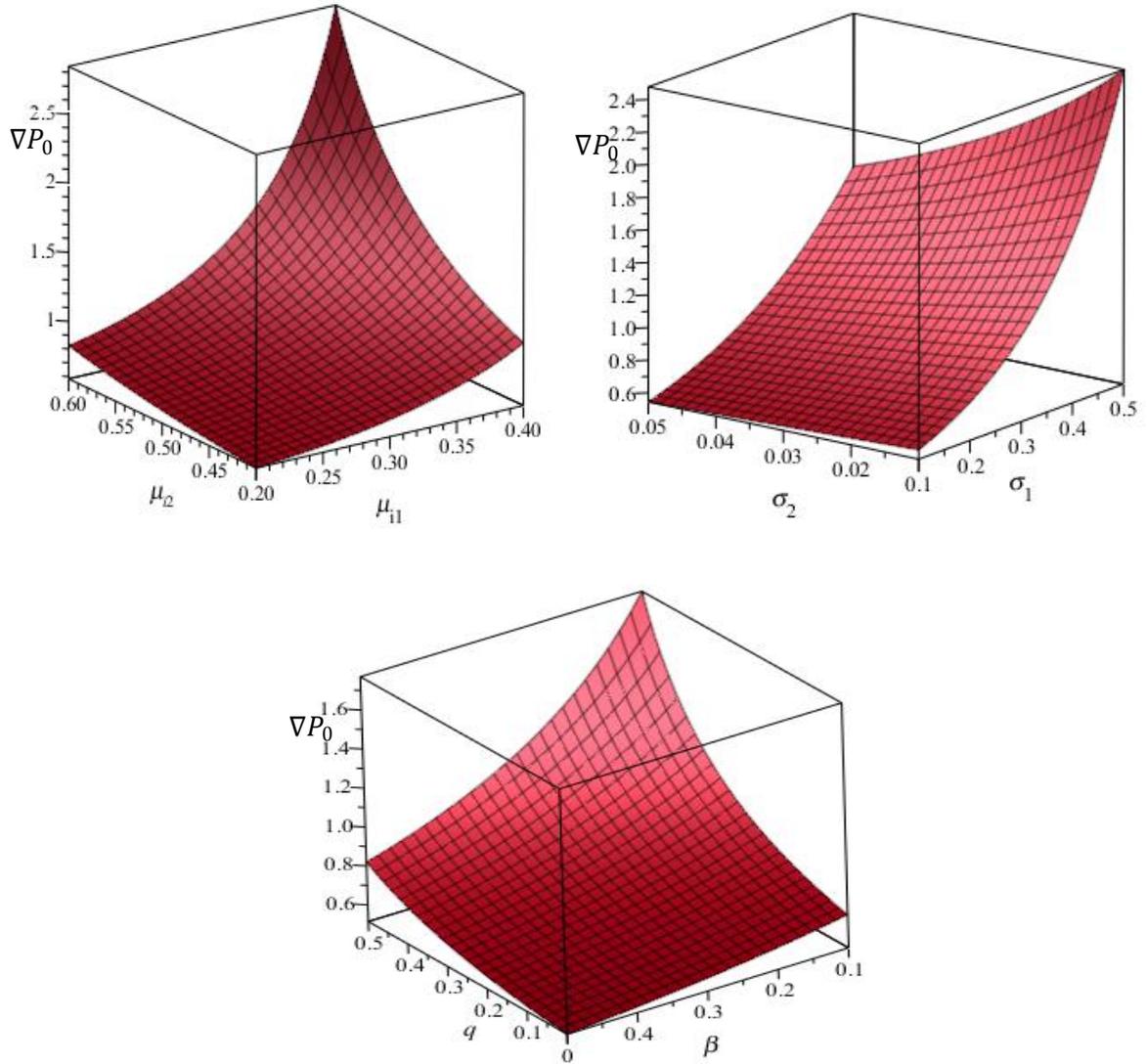

Fig. 2 Influence on phase shifts due to head-on collision for two-soliton (a) $\mu_{i1}$ and $\mu_{i2}$ (b) $\sigma_1$ and $\sigma_2$ and (c) $q$ and $\beta$ considering typical values as of Fig.1.

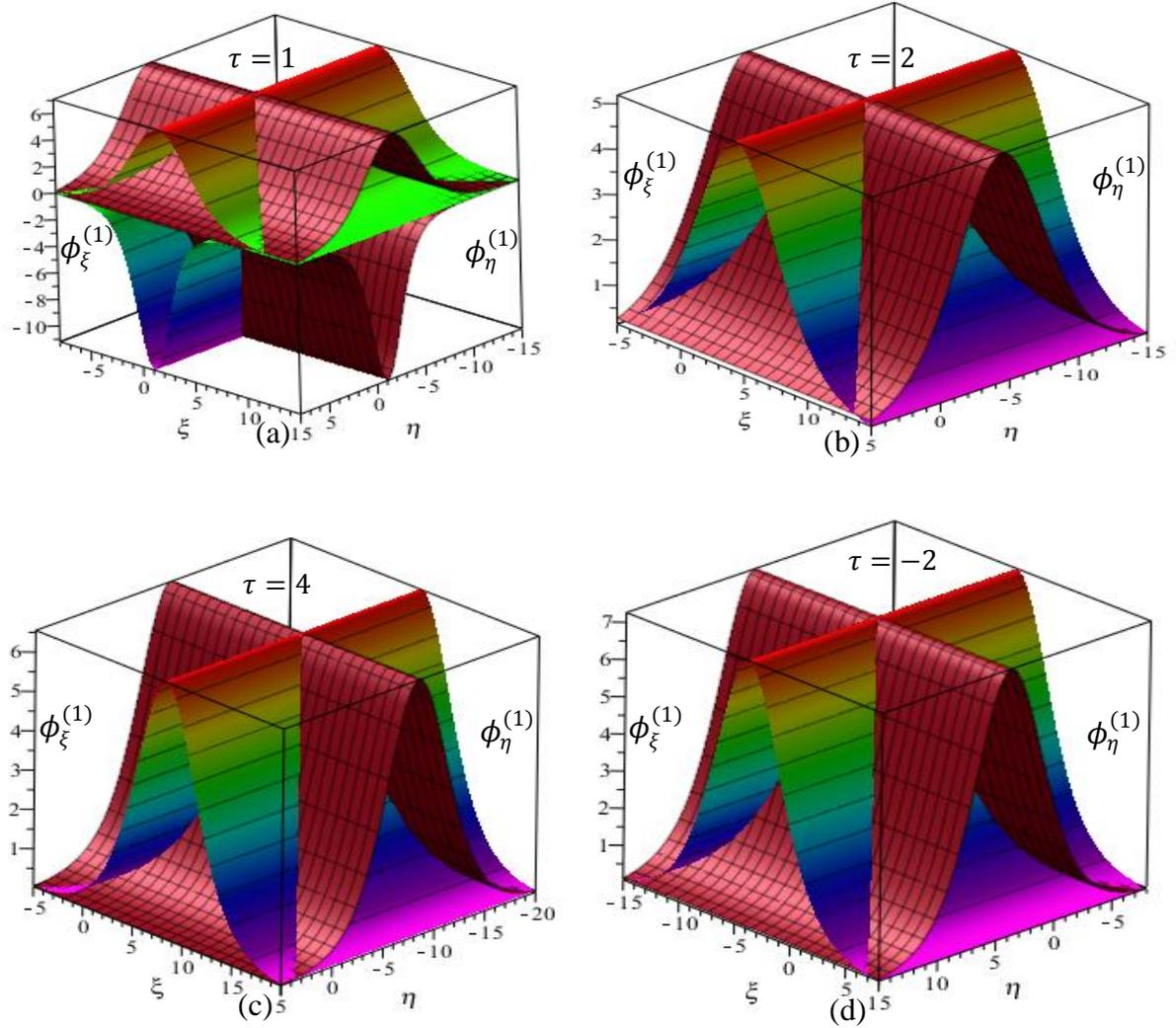

Fig. 3 Electrostatic potential profiles due to head-on collisions between $\phi_\xi^{(1)}(\xi,\tau)$ and $\phi_\eta^{(1)}(\eta,\tau)$ for one-soliton taking (a) $\mu_{i1} = 0.2$ (compressive), $\mu_{i1} = 0.78$ (rarefactive), $\mu_{i2} = 0.41$ $\sigma_1 = 0.2$, $\sigma_2 = 0.05$, $\beta = 0.3$ and $q = 0.25$, (b) $\mu_{i1} = 0.2$, $\mu_{i2} = 0.41$, $\sigma_1 = 0.1$, $\sigma_2 = 0.05$, $\beta = 0.3$ and $q = 0.25$, (c) same values of (b) but $\sigma_2 = 0.01$, and (d) same values of (c) but $\mu_{i2} = 0.5$.

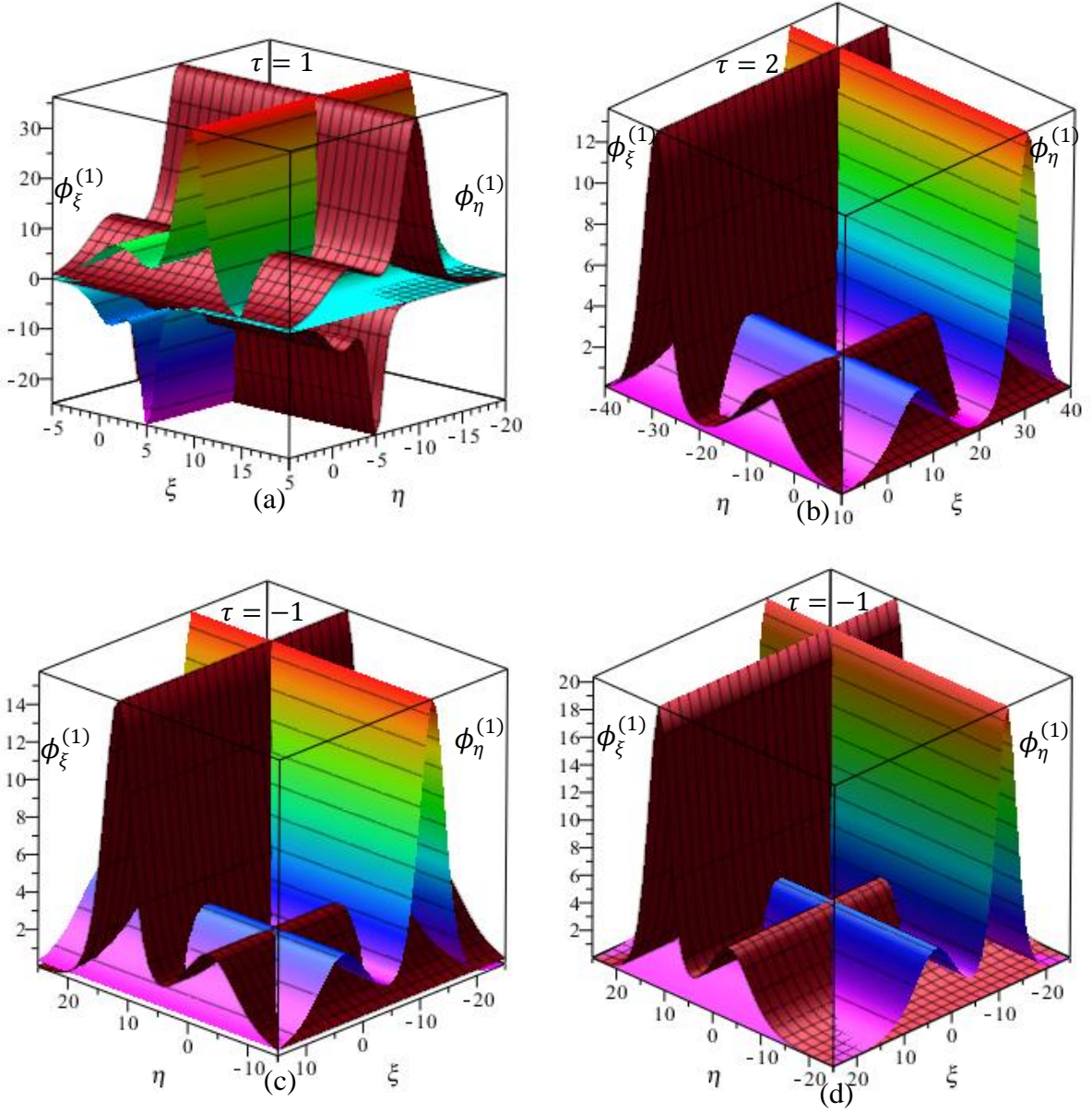

Fig. 4 Electrostatic potential profiles due to head-on collisions between $\phi_\xi^{(1)}(\xi,\tau)$ and $\phi_\eta^{(1)}(\eta,\tau)$ for two-soliton taking (a) $\mu_{i1} = 0.1$ (compressive), $\mu_{i1} = 0.78$ (rarefactive), $\mu_{i2} = 0.41$ $\sigma_1 = 0.5$, $\sigma_2 = 0.05$, $\beta = 0.3$ and $q = 0.25$, (b) $\mu_{i1} = 0.1$, $\mu_{i2} = 0.41$, $\sigma_1 = 0.1$, $\sigma_2 = 0.05$, $\beta = 0.5$ and $q = 0.25$, (c) same values of (b) except $q = 0.5$, and (d) same values of (b) except $\sigma_2 = 0.01$.

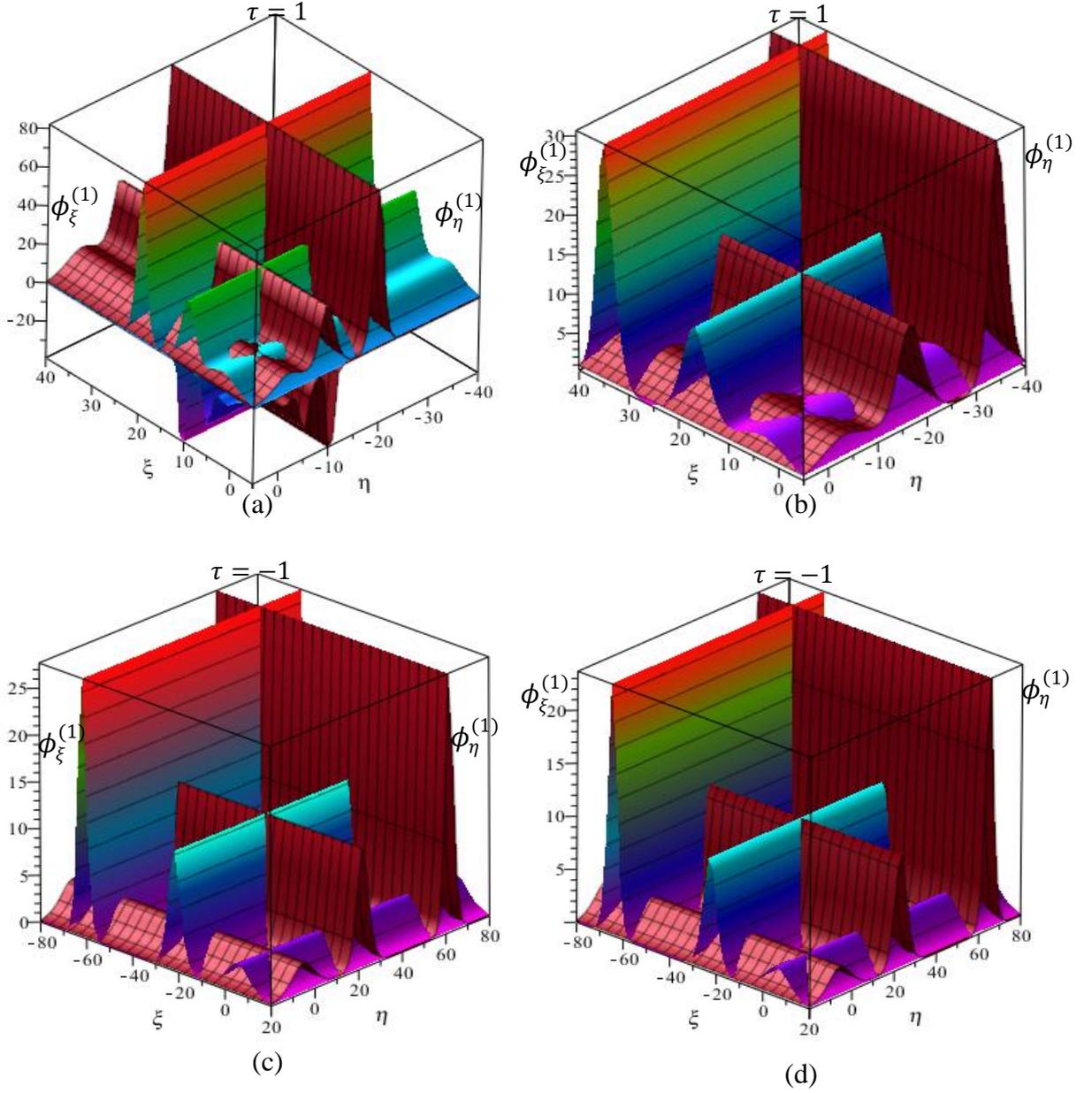

Fig. 5 Electrostatic potential profiles due to head-on collisions between $\phi_\xi^{(1)}(\xi,\tau)$ and $\phi_\eta^{(1)}(\eta,\tau)$ for three-soliton taking (a) $\mu_{i1} = 0.1$ (compressive), $\mu_{i1} = 0.78$ (rarefactive), $\mu_{i2} = 0.41$ $\sigma_1 = 0.5$, $\sigma_2 = 0.05$, $\beta = 0.3$ and $q = 0.25$, (b) $\mu_{i1} = 0.1$, $\mu_{i2} = 0.41$, $\sigma_1 = 0.1$, $\sigma_2 = 0.05$, $\beta = 0.5$ and $q = 0.25$, (c) same values of (b) but $q = 0.15$, and (d) same values of (b) but $\mu_{i2} = 0.31$.